# Using Apriori with WEKA for Frequent Pattern Mining

Paresh Tanna[#1], Dr. Yogesh Ghodasara[*2]

[1*] *MCA Department, RK. University, Rajkot, Gujarat, INDIA*
[2] *College of Information Tech., Anand Agriculture University, Anand, Gujarat, INDIA*

*Abstract*— Knowledge exploration from the large set of data, generated as a result of the various data processing activities due to data mining only. Frequent Pattern Mining is a very important undertaking in data mining. Apriori approach applied to generate frequent item set generally espouse candidate generation and pruning techniques for the satisfaction of the desired objective. This paper shows how the different approaches achieve the objective of frequent mining along with the complexities required to perform the job. This paper demonstrates the use of WEKA tool for association rule mining using Apriori algorithm.

## I. INTRODUCTION

Due to extensive mechanization and due to affordable storage facilities, there is an massive prosperity of information embedded in huge database belonging to different enterprise. The main idea of data mining is to find effective ways to combine the computer's power to process data with human eye's ability to detect patterns. The techniques of data-mining are designed for and work best with, large data sets[5].

Today, the processor is having speed that is underutilized due to improper localization of the various parameters if these parameters would be properly localized than the performance of the system can be improved a lot. This can be done using several cache conscious mechanisms that are going to help in optimal use of the resources for better outcome. Here we have discussed the mechanism for WEKA to use Apriori algorithm.

## II. THE APRIORI ALGORITHM

The Apriori Algorithm is an influential algorithm for mining frequent itemsets for boolean association rules. Some key concepts for Apriori algorithm are :

- Frequent Itemsets: The sets of item which has minimum support (denoted by Li for ith-Itemset).
- Apriori Property: Any subset of frequent itemset must be frequent.
- Join Operation: To find Lk , a set of candidate k-itemsets is generated by joining Lk-1 with itself.

Very first algorithm proposed for association rules mining was the Apriori for frequent itemset mining[1]. The most popular algorithm for pattern mining is without a doubt Apriori. It is designed to be applied on a transaction database to discover patterns in transactions made by customers in stores. But it can also be applied in several other applications. A transaction is defined a set of distinct items (symbols). Apriori takes as input (1) a minsup threshold set by the user and (2) a transaction database containing a set of transactions. Apriori outputs all frequent itemsets, i.e. groups of items shared by no less than minsup transactions in the input database. For example, consider the following transaction database containing four transactions. Given a minsup of two transactions, frequent itemsets are "bread, butter", "bread milk", "bread", "milk" and "butter".[1,6]

T1: bread, butter, spinach
T2: butter, salmon
T3: bread, milk, butter
T4: cereal, bread, milk

The Apriori algorithm employs the downward closure property if an item set is not frequent, any superset of it cannot be frequent either. The Apriori algorithm performs a breadth-first search in the search space by generating candidate k+1-itemsets from frequent k itemsets[1].

The frequency of an item set is computed by counting its occurrence in each transaction. Apriori is an significant algorithm for mining frequent itemsets for Boolean association rules. Since the Algorithm uses prior knowledge of frequent item set it has been given the name Apriori. Apriori is an iterative level wise search Algorithm, where k- itemsets are used to explore (k+1)-itemsets. First, the set of frequents 1-itemsets is found[1].

This set is denoted by L1. L1 is used to find L2 , the set of frequent 2-itemsets , which is used to find L3 and so on , until no more frequent k-itemsets can be found. The finding of each Lk requires one full scan of database[1,6].

There are two steps for understanding that how Lk-1 is used to find Lk:-

1) The join step : To find Lk , a set of candidate k-itemsets is generated by joining Lk-1 with itself . This set of candidates is denoted Ck.
2) The prune step: Ck is a superset of Lk , that is , its members may or may not be frequent , but all of the frequent k-itemsets are included in Ck .

A scan of the database to determine the count of each candidate in Ck would result in the determination of Lk. Ck, however, can be huge, and so this could involve heavy computation.

To reduce the size of Ck , the Apriori property is used as follows.

i. Any (k-1)-item set that is not frequent cannot be a subset of frequent k-item set.

ii. Hence, if (k-1) subset of a candidate k item set is not in Lk-1 then the candidate cannot be frequent either and so can be removed from C.

Based on the Apriori property that all subsets of a frequent itemset must also be frequent, we can determine that four latter candidates cannot possibly be frequent. How ?

For example, let's take {I1, I2, I3}. The 2-item subsets of it are {I1, I2}, {I1, I3} & {I2, I3}. Since all 2-item subsets of {I1, I2, I3} are members of L2, We will keep {I1, I2, I3} in C3.

Let's take another example of {I2, I3, I5} which shows how the pruning is performed. The 2-item subsets are {I2, I3}, {I2, I5} & {I3,I5}.

BUT, {I3, I5} is not a member of L2 and hence it is not frequent violating Apriori Property. Thus We will have to remove {I2, I3, I5} from C3.

Therefore, C3 = {{I1, I2, I3}, {I1, I2, I5}} after checking for all members of result of Join operation for Pruning.

**Apriori algorithm pseudo code:**

procedure **Apriori** (T, *minSupport*)

{ //T is the database and *minSupport* is the minimum support

    L1= {frequent items};

    **for** (k= 2; Lk-1 !=∅; k++)

    {

        Ck= candidates generated from Lk-1

        //that is cartesian product Lk-1 x Lk-1 and

        //eliminating any k-1 size itemset that is not

        //frequent

        **for each** transaction **t** in database **do**

        {

        #increment the count of all candidates in Ck

        that are contained in t

        Lk = candidates in Ck with *minSupport*

    }//end for each

    }//end for

**return** ;

}

Association rule generation is usually split up into two separate steps:

1. First, minimum support is applied to find all *frequent itemsets* in a database.

2. Second, these frequent itemsets and the minimum confidence constraint are used to form rules.

While the second step is straight forward, the first step needs more attention. Finding all frequent itemsets in a database is difficult since it involves searching all possible itemsets (item combinations).

The set of possible itemsets is the power set over I and has size $2^n - 1$ (excluding the empty set which is not a valid itemset). Although the size of the powerset grows exponentially in the number of items n in I, efficient search is possible using the downward-closure property of support (also called anti-monotonicity) which guarantees that for a frequent itemset, all its subsets are also frequent and thus for an infrequent itemset, all its supersets must also be infrequent. Exploiting this property, efficient algorithm (e.g., Apriori) can find all frequent itemsets. As is common in association rule mining, given a set of itemsets (for instance, sets of retail transactions, each listing individual items purchased), the algorithm attempts to find subsets which are common to at least a minimum number min_sup of the itemsets. Apriori uses a "bottom up" approach, where frequent subsets are extended one item at a time (a step known as candidate generation), and groups of candidates are tested against the data[6].

The algorithm terminates when no further successful extensions are found. Apriori uses breadth-first search and a tree structure to count candidate item sets efficiently. It generates candidate item sets of length *k* from item sets of length *k* − 1. Then it prunes the candidates which have an infrequent sub pattern.

According to the downward closure lemma, the candidate set contains all frequent *k*-length item sets. After that, it scans the transaction database to determine frequent item sets among the candidates.

Apriori, while historically significant, suffers from a number of inefficiencies or trade-offs, which have spawned other algorithms. Candidate generation generates large numbers of subsets (the algorithm attempts to load up the candidate set with as many as possible before each scan).

Bottom-up subset exploration (essentially a breadth-first traversal of the subset lattice) finds any maximal subset S only after all $2^{|S|} - 1$ of its proper subsets.

### III. SAMPLE USAGE OF ASSOCIATION RULE MINING IN WEKA:

For our test we shall consider 15 transactions that have made for a shopping center. Each transaction has specific list of items. Here we have demonstrated use of Apriori algorithm for association rule mining using WEKA[8]. The ARFF file presented bellow contains information regarding each transaction's items detail.

Transaction table:

| Trans ID | Items |
|---|---|
| 1 | A,B,C,D,G,H |
| 2 | A,B,C,D,E,F,H |
| 3 | B,C,D,E,H |
| 4 | B,E,G,H |
| 5 | A,B,D,E,G,H |
| 6 | A,C,F,G,H |
| 7 | B,D,E,G,H |
| 8 | A,C,D,E,G,H |
| 9 | B,C,D,E,H |
| 10 | A,C,E,F,H |
| 11 | C,E,H |
| 12 | A,D,E,F,H |
| 13 | B,C,E,F,H |
| 14 | A,B,C,F,H |
| 15 | A,B,E,F,H |

TEST_ITEM_TRANS.arff

@relation TEST_ITEM_TRANS
@attribute A {TRUE, FALSE}
@attribute B {TRUE, FALSE}
@attribute C {TRUE, FALSE}
@attribute D {TRUE, FALSE}
@attribute E {TRUE, FALSE}
@attribute F {TRUE, FALSE}
@attribute G {TRUE, FALSE}
@attribute H {TRUE, FALSE}
@data

TRUE,TRUE,TRUE,TRUE,FALSE,FALSE,TRUE,TRUE
TRUE,TRUE,TRUE,TRUE,TRUE,TRUE,FALSE,TRUE
FALSE,TRUE,TRUE,TRUE,FALSE,FALSE,FALSE,TRUE
FALSE,TRUE,FALSE,FALSE,TRUE,FALSE,TRUE,TRUE
TRUE,TRUE,FALSE,TRUE,TRUE,FALSE,TRUE,TRUE
TRUE,FALSE,TRUE,FALSE,FALSE,TRUE,TRUE,TRUE
FALSE,TRUE,FALSE,TRUE,TRUE,FALSE,TRUE,TRUE
TRUE,FALSE,TRUE,TRUE,TRUE,FALSE,TRUE,TRUE
FALSE,TRUE,TRUE,TRUE,TRUE,FALSE,FALSE,TRUE
TRUE,FALSE,TRUE,FALSE,TRUE,TRUE,FALSE,TRUE
FALSE,FALSE,TRUE,FALSE,TRUE,FALSE,FALSE,TRUE
TRUE,FALSE,FALSE,TRUE,TRUE,TRUE,FALSE,TRUE
FALSE,TRUE,TRUE,FALSE,TRUE,TRUE,FALSE,TRUE
TRUE,TRUE,TRUE,FALSE,FALSE,TRUE,FALSE,TRUE
TRUE,TRUE,FALSE,FALSE,TRUE,TRUE,FALSE,TRUE

---

Using the Apriori Algorithm we want to find the association rules that have **minSupport=50%** and minimum confidence=50%. We will do this using WEKA GUI.

After we launch the WEKA application and open the *TEST_ITEM_TRANS.arff* file as shown in figure 1. Then we move to the **Associate** tab and we set up the configuration as shown in figure 2.

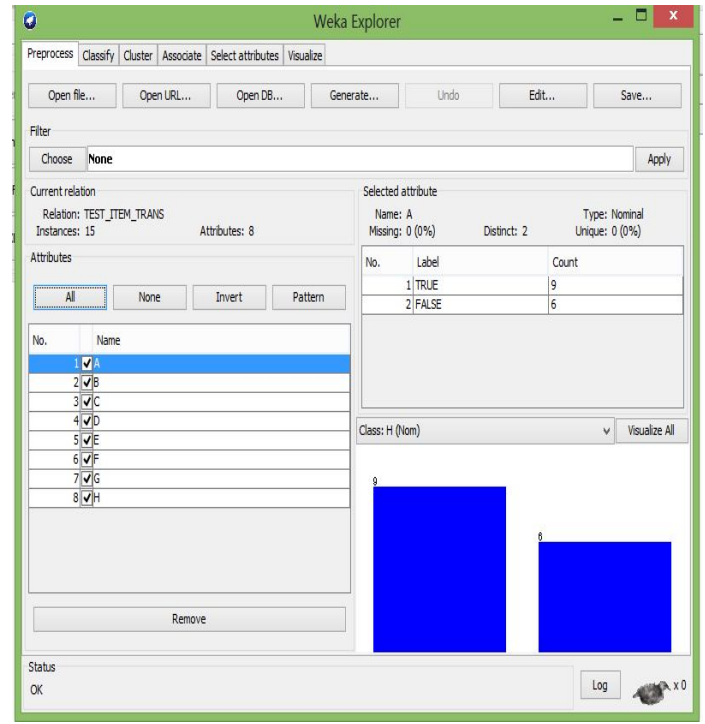

Figure 1: Opening file in Weka Explorer

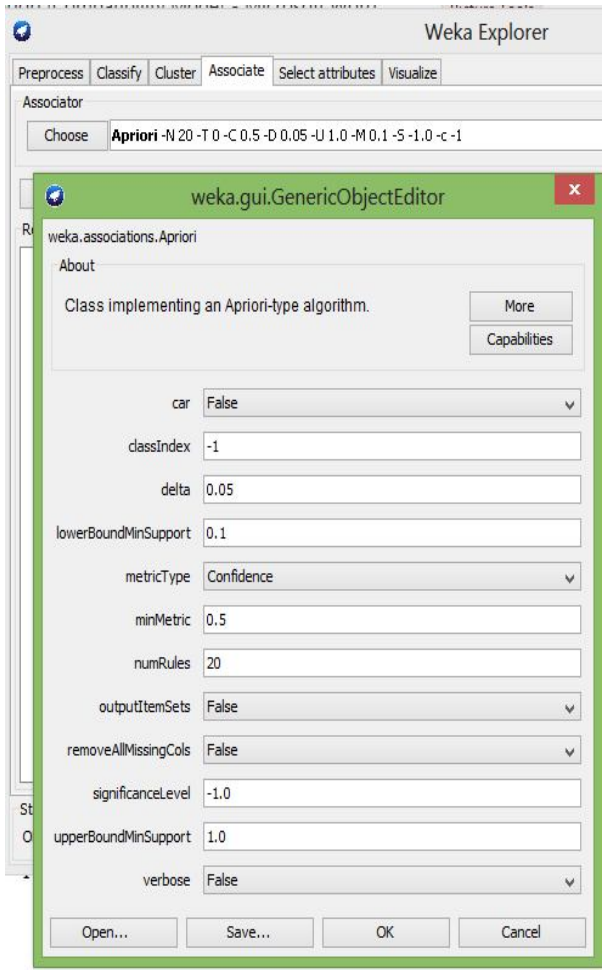

Figure 2: Attribute configuration in Weka Explorer

After the algorithm is finished, we get the following results:

=== Run information ===

Scheme:     weka.associations.Apriori -N 20 -T 0 -C 0.5 -D 0.05 -U 1.0 -M 0.1 -S -1.0 -c -1
Relation:    TEST_ITEM_TRANS
Instances:   15
Attributes:  8
     A
     B
     C
     D
     E
     F
     G
     H

=== Associator model (full training set) ===

Apriori
=======
Minimum support: 0.5 (7 instances)
Minimum metric <confidence>: 0.5
Number of cycles performed: 10

Generated sets of large itemsets:

Size of set of large itemsets L(1): 10

Size of set of large itemsets L(2): 12

Size of set of large itemsets L(3): 3

Best rules found:

 1. E=TRUE 11 ==> H=TRUE 11    conf:(1)
 2. B=TRUE 10 ==> H=TRUE 10    conf:(1)
 3. C=TRUE 10 ==> H=TRUE 10    conf:(1)
 4. A=TRUE 9 ==> H=TRUE 9    conf:(1)
 5. G=FALSE 9 ==> H=TRUE 9    conf:(1)
 6. D=TRUE 8 ==> H=TRUE 8    conf:(1)
 7. F=FALSE 8 ==> H=TRUE 8    conf:(1)
 8. D=FALSE 7 ==> H=TRUE 7    conf:(1)
 9. F=TRUE 7 ==> H=TRUE 7    conf:(1)
10. B=TRUE E=TRUE 7 ==> H=TRUE 7    conf:(1)
11. C=TRUE G=FALSE 7 ==> H=TRUE 7    conf:(1)
12. E=TRUE G=FALSE 7 ==> H=TRUE 7    conf:(1)
13. G=FALSE 9 ==> C=TRUE 7    conf:(0.78)
14. G=FALSE 9 ==> E=TRUE 7    conf:(0.78)
15. G=FALSE H=TRUE 9 ==> C=TRUE 7    conf:(0.78)
16. G=FALSE 9 ==> C=TRUE H=TRUE 7    conf:(0.78)
17. G=FALSE H=TRUE 9 ==> E=TRUE 7    conf:(0.78)
18. G=FALSE 9 ==> E=TRUE H=TRUE 7    conf:(0.78)
19. H=TRUE 15 ==> E=TRUE 11    conf:(0.73)
20. B=TRUE 10 ==> E=TRUE 7    conf:(0.7)

## IV.  APRIORI GOAL AND APPLICATION

**Goal:** finding inherent regularities in data
- What products were often purchased together?— Beer and diapers?
- What are the subsequent purchases after buying a PC?
- What kinds of DNA are sensitive to this new drug?
- Can we automatically classify Web documents?

**Applications:**
- Basket data analysis, cross-marketing, catalog design, sale campaign analysis, Web log (click stream) analysis, and DNA sequence analysis[4,6].

## V. EXPERIMENTAL RESULTS

We ran experiments with Apriori program on a dataset, which so that the advantages and disadvantages of the approach and the different optimizations can be observed.

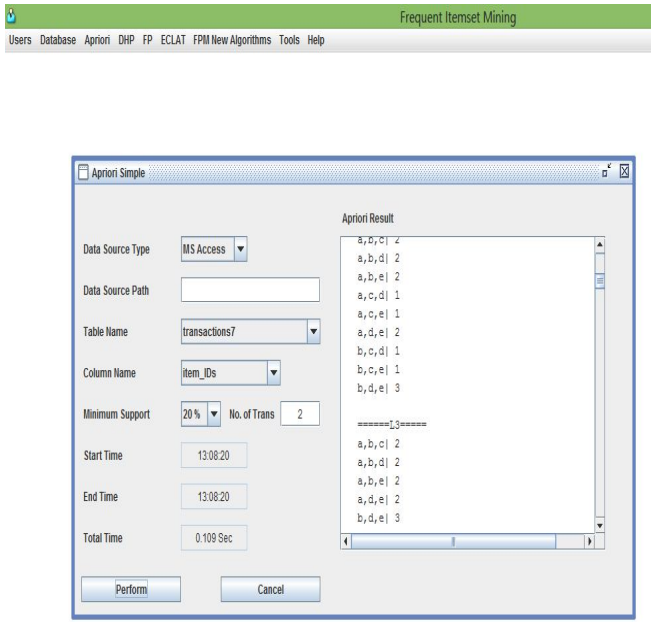

Figure 3: Apriori algorithm implementation result

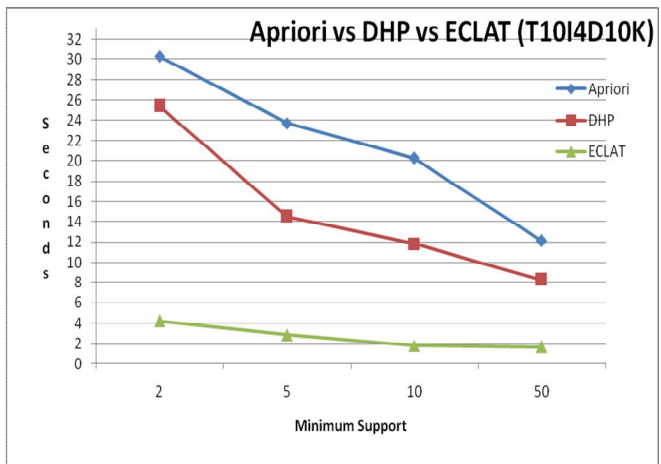

Figure 4: Apriori Vs DHP Vs ECLAT (T10I4D10K)[1,2,3,7]

## VI. CONCLUSION

Apriori is the simplest algorithm which is used for mining of frequent patterns from the transaction database. The purpose of reducing the number of scans of database to extract frequent item set will be resolved in future due to our work is in progress for the same. We have tried to implement the Apriori algorithm for sufficient research work and also we have utilized WEKA for referring the process of association rule mining.